\documentclass[useAMS,usenatbib]{mn2e}
\usepackage{graphicx}
\usepackage{enumerate}
\usepackage{epsfig}

\title[Accretion disc variability in BHXRB hard states]{Accretion disc variability in the hard state of black hole X-ray binaries}
\author[T. Wilkinson \& P. Uttley]{Tony Wilkinson$^{1}$\thanks{E-mail:
t.r.wilkinson@phys.soton.ac.uk} and Philip Uttley$^{1}$\\
$^{1}$School of Physics and Astronomy, University of Southampton, Southampton SO17 1BJ}

\begin{document}

\date{}

\pagerange{\pageref{firstpage}--\pageref{lastpage}} \pubyear{2008}

\maketitle

\label{firstpage}

\begin{abstract}
{\it XMM-Newton} X-ray spectra of the hard state Black Hole X-Ray Binaries (BHXRBs) SWIFT~J1753.5-0127 and GX~339-4 show evidence for accretion disc blackbody emission, in addition to hard power-laws.  The soft and hard band Power-Spectral Densities (PSDs) of these sources demonstrate variability over a wide range of time-scales. However, on time-scales of tens of seconds, corresponding to the putative low-frequency Lorentzian in the PSD, there is additional power in the soft band. To interpret this behaviour, we introduce a new spectral analysis technique, the `covariance spectrum', to disentangle the contribution of the X-ray spectral components to variations on different time-scales.  We use this technique to show that the disc blackbody component varies on all time-scales, but varies more, relative to the power-law, on longer time-scales. This behaviour explains the additional long-term variability seen in the soft band.  Comparison of the blackbody and iron line normalisations seen in the covariance spectra in GX~339-4 implies that the short-term blackbody variations are driven by thermal reprocessing of the power-law continuum absorbed by the disc. However, since the amplitude of variable reflection is the same on long and short time-scales, we rule out reprocessing as the cause of the enhanced disc variability on long time-scales.  Therefore we conclude that the long-time-scale blackbody variations are caused by instabilities in the disc itself, in contrast to the stable discs seen in BHXRB soft states.  Our results provide the first observational evidence that the low-frequency Lorentzian feature present in the PSD is produced by the accretion disc.
\end{abstract}

\begin{keywords}

\end{keywords}

\section{Introduction}
X-ray spectra of BHXRBs show evidence for a two-phase structure to the accretion flow, an optically thick, geometrically thin accretion disc \citep{Shakura1973} giving rise to a blackbody component in the X-ray spectrum, and a hot optically-thin component, modelled as a power law. The relative strengths of these two components define the appearance of different `states' (e.g. \citealt{Homan2001}).  In the hard state, which we focus on in this paper, the power-law emission dominates the total luminosity.  It has been suggested that the power-law is produced by an inner, optically-thin Advection Dominated Accretion Flow (ADAF) (e.g. \citealt{Esin1997}), which replaces the inner optically thick disc at low accretion rates, and extends down to the Innermost Stable Circular Orbit (ISCO). This implies that the optically thick disc is truncated at some transition radius. Alternatively, the optically thick and optically thin components may co-exist over some range of radii, e.g. if the thin disc is sandwiched by a hot flow or corona \citep{Witt1997,Churazov2001}.  The corona may in turn evaporate the innermost regions of the disc at low accretion rates (e.g. \citealt{Meyer-Hofmeister2003,Mayer2007}), so that the optically thin flow is radially separated from the optically thick disc as for the ADAF model. It has also been suggested \citep{Malzac2007} that a cold thin accretion disc extends close to the black hole in the hard-state. In this model, most of the accretion power is transported away from the disc to power a strong outflowing corona and jet. Since the ADAF and corona perform similar roles in that they upscatter soft photons to produce the observed power-law, we shall henceforth refer to both components interchangeably as the corona, without necessarily favouring either picture.

In the hard state, the variability of the power-law continuum offers further clues to the structure of the accretion flow.  Studies of the timing properties of hard state BHXRBs show that the frequencies of broad Lorentzian features in their Power-Spectral Density functions (PSDs) correlate with the strength of the reflection features as well as the steepness of the power law continuum \citep{Revnivtsev2001,Gilfanov1999}. These correlations can naturally be explained if the Lorentzian frequencies correspond to a characteristic timescale at the disc truncation radius, e.g. the viscous time-scale, so that as the truncation radius increases the Lorentzian frequency decreases, with disc reflection and Compton cooling of the optically thin hot flow by disc photons decreasing accordingly.  In this picture, the truncation radius of the thin disc acts to generate the signals of the lowest-frequency Lorentzian in the PSD, while the highest-frequency Lorentzians may be generated at the innermost radius of the hot inner coronal flow, i.e., at the ISCO \citep{Done2007}.

Regardless of whether the corona is radially or vertically separated from the thin disc, photons upscattered by the corona should interact with the disc. This interaction gives rise to reflection features in the X-ray spectrum, including fluorescent iron line emission and a reflection continuum due to Compton scattering off the disc material. An often-neglected consideration is that a significant fraction of the photons interacting with the disc are absorbed and the disc heated in a process known as thermal reprocessing. Provided that the disc subtends at least a moderate solid angle as seen by the corona, this effect should be particularly significant in the hard state, where the coronal power law continuum dominates the total luminosity. When the power-law luminosity impinging on the disc is high compared to the disc luminosity due to internal heating, then a significant fraction of the disc blackbody emission should be reprocessed and will therefore track variations of the power law continuum.  

The anticipated correlated variations of the blackbody and power-law emission can be studied using variability spectra, e.g. the rms spectrum, which show only the variable components of the spectrum \citep{Vaughan2003}.  If the geometry is such that the observed power-law produces reprocessed blackbody emission by X-ray heating the disc, then both power-law and blackbody components should appear together in the variability spectra.  Furthermore, by selecting different time ranges covered by these variability spectra (i.e., analogous to the method of Fourier-resolved spectrosopy, \citealt{Revnivtsev1999}), it is possible to determine whether the low-frequency part of the PSD has a different origin to the high-frequency part in terms of the contributions of blackbody and power-law components.  If the optically thick disc does drive the low-frequency Lorentzian, we predict that the blackbody component should be stronger in the corresponding variability spectrum.

In this work, we examine the variability spectra of two hard state black hole X-ray binaries, SWIFT J1753.5-0127 and GX 339-4, which have good {\it XMM-Newton} data for studying simultaneous variations of disc and power-law.
Previous analyses of time-averaged spectra for these {\it XMM-Newton} observations have shown the existence of the expected blackbody components, together with relativistically broadened reflection features, which have been used to argue that the disc is truncated at significantly smaller radii than previously thought, perhaps only a few gravitational radii \citep{Miller2006a,Miller2006b}. However, see \citealt{Gierlinski2008,Hiemstra2009,Done2007} for arguments in favour of disc truncation at larger radii. In the following Section, we describe the observations and data reduction. In Section~\ref{anres} we show the soft and hard-band PSDs obtained from the data, and present a technique to produce a type of rms spectrum, the `covariance spectrum' which we use to identify the variable spectral components for each hard state source.  In particular we show that, although both power-law and disc blackbody emission are correlated, as expected from thermal reprocessing, the disc is relatively more variable than the power-law on longer time-scales (corresponding to the low-frequency Lorentzian), contrary to what we would expect from simple reprocessing models.  In Section~\ref{discuss} we discuss the interpretation of our results and present further analysis to suggest that the low-frequency Lorentzian corresponds to fluctuations intrinsic to the disc.  A summary of our conclusions is given in Section~\ref{conc}.

\section{Observations and Data Reduction}
\label{obsred}
SWIFT J1753.5-0127 was observed during revolution 1152 for 42 ks by \textit{XMM-Newton} EPIC-pn on 2006 March 24 in \textit{pn-timing} mode using the medium optical filter. The events list was screened using the perl script \textsc{xmmclean}\footnote{http://lheawww.gsfc.nasa.gov/$\sim$kaa/xselect/xmmclean} to select only events with FLAG=0 and PATTERN $\leq$ 4. Examination of light curves showed no evidence for background flaring. Using SAS Version 7.0, \textsc{evselect} was used to extract the mean spectrum, following the procedure of \citet{Miller2006b}, using event positions between 20 and 56 in RAWX and using the full RAWY range. The SAS tasks \textsc{arfgen} and \textsc{rmfgen} were used to generate the Ancillary Response File (ARF) and Redistribution Matrix File (RMF).

GX 339-4 was observed by \textit{XMM-Newton} on 2004 March 16 during revolutions 782 and 783, again in \textit{pn-timing} mode using the medium optical filter. The SAS command \textsc{evselect} was used to filter the events on TIME to avoid background flaring, producing a total combined exposure of 127~ks. Data reduction was very similar to that for SWIFT J1753.5-0127, using SAS version 7.0, but care was taken to avoid pile-up. Successive columns from the centre of the image were excised in RAWX and spectra extracted until no discernable difference in spectral shape could be identified between successive selections on RAWX, implying that pile-up is no longer significant. The EPIC-pn data was deemed free of pile-up using extraction regions in RAWX from columns 30 to 36 and 40 to 46 (i.e., columns 37 to 39 inclusive were excised).  Background spectra were selected in RAWX from columns 10 to 18 over the full RAWY range. To generate an appropriate ARF for the data made in this way, it was necessary to use \textsc{arfgen} to generate an ARF for the full region in RAWX from columns 30 to 46 and generate a second ARF from the spectrum of the excluded region (RAWX columns 37 to 39) and then subtract the latter from the former using the command \textsc{addarf}\footnote{http://xmm.vilspa.esa.es/external/xmm\_user\_support\\/documentation/sas\_usg/USG/node63.html}.

For both sources, the \textsc{ftool} \textsc{grppha} was used to scale the background spectra and to ensure that a minimum of 20 counts were in each bin for $\chi^2$ fitting. The EPIC-pn covers the energy range from 0.2-10~keV, but the EPIC Calibration Status Document\footnote{http://xmm2.esac.esa.int/external/xmm\_sw\_cal/calib/index.shtml} recommends restricting the fit to energies greater than 0.5~keV. To be conservative, our EPIC-pn spectral fits were restricted to the range 0.7 to 10.0 keV.  

We extracted data from \textit{RXTE} observations which were contemporaneous with the \textit{XMM-Newton} observations. We used the high time and spectral-resolution PCA Event mode data to extract mean spectra and make rms and covariance spectra using the same time binning as the EPIC-pn data (see Section~\ref{method}). Throughout this work we use \textit{RXTE} PCA data in the 3-25 keV range, and fit these together with the corresponding EPIC-pn spectra, tying all fit parameters together but allowing \textit{RXTE} spectral fits to be offset by a constant factor with respect to the simultaneous EPIC-pn fits, to allow for the difference in flux calibration between the PCA and EPIC-pn instruments, and also for the fact that slight flux differences may result from the fact that the {\it RXTE} observations covered shorter intervals than the {\it XMM-Newton} data.  A 1~per cent systematic error was assumed in all spectral fits to account for uncertainties in instrumental response and cross-calibration. The signal to noise of \textit{RXTE} HEXTE data was not sufficient to perform the covariance analysis we describe in section 3.2. 

\section[]{Analysis and Results}
\label{anres}
\subsection{Power-spectral densities}
\begin{figure*}
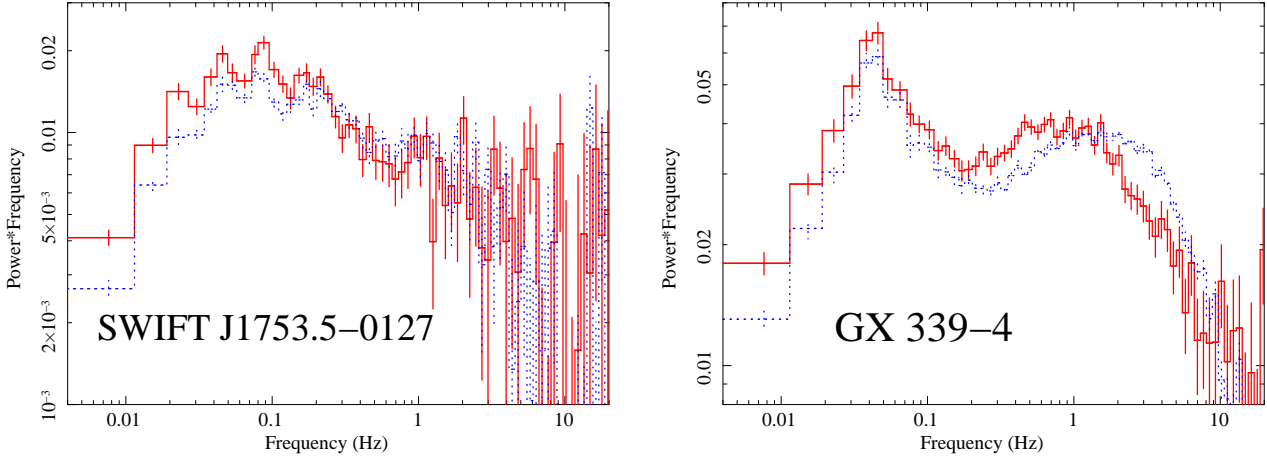

\centering
\includegraphics[width=60mm,angle=-90]{swiftpsd.ps}
\hspace{0.5cm}
\includegraphics[width=60mm,angle=-90]{gx339psd.ps}
\caption{Soft and hard band PSDs for SWIFT~J1753.5-0127 and GX~339-4.  The solid line denotes the 0.5-1~keV PSD, while the dotted line denotes the 2-10~keV PSD.  The Poisson noise component has been subtracted from all PSDs.}
\label{psds}
\end{figure*}

Before describing the spectral analysis method and results, we first examine the timing properties of each source in soft and hard spectral bands using the power-spectral density function (PSD).  For each source, we used the EPIC-pn data to generate light curves with 1~ms time binning in two energy bands: 0.5-1~keV (soft) and 2-10~keV (hard).  Only complete segments of 131~s duration\footnote{Specifically, containing 131072 time bins.} were used to construct the PSDs, segments with gaps (which can be common at high count rates in timing mode) were skipped over. The resulting Poisson-noise-subtracted PSDs are shown in figure~\ref{psds}.  It is clear from the figure that the PSDs are significantly different in shape between the two bands.  In GX~339-4, the hard-band PSD shows two broad components, with the higher-frequency component appearing to be shifted to even higher frequencies in the hard band. In SWIFT~J1753.5-0127, soft and hard PSDs appear, within the noise, to overlap more closely at higher frequencies.  However, in both sources the soft band PSD shows relatively larger low-frequency power compared to the PSD components at frequencies above $\sim~0.2$~Hz.  This $\sim~20\%$ extra low-frequency power at lower X-ray energies may be associated with the soft excess emission, e.g. if the disc is varying more than the power-law on longer time-scales. Alternatively, there could be power-law spectral variability that applies only on longer time-scales and causes the soft band variability amplitude to be enhanced, i.e., due to steepening of the power-law spectral slope on longer time-scales.  To determine the origin of the extra low-frequency power in the soft band, we must carry out a spectral analysis of the variations on different time-scales, the methodology for which we describe in the next section.

\subsection[]{Methodology: rms and covariance spectra}
\label{method}
When fitting models to X-ray spectra, it is typical to use mean X-ray spectra that only describe the time-averaged spectral shape of a source. Such fits say nothing about the way the different spectral components such as disc blackbody and power-law vary with respect to each other in time. By looking at the absolute amplitude of variations in count rate as a function of energy we can construct `variability spectra' which pick out only the time-varying components.  One technique is to construct Fourier-frequency resolved spectra, obtaining a PSD for each individual energy channel and integrating the PSD over a given Fourier-frequency range in order to measure the variance in that channel, which is used to obtain the rms and so construct the spectrum due to components which vary over that frequency range (e.g. \citealt{Revnivtsev1999}).  This approach is attractive in that it allows the user to pick out complex patterns of spectral variability where components have different time-scales of variation, as may be implied by the soft and hard band PSDs shown in figure~\ref{psds}.  Here we will consider only two time-scales of variability, corresponding roughly to the high and low-frequency parts of the PSD which show significant relative differences between the soft and hard bands.  To approximate the more complex Fourier-resolved approach we will measure over two time-scale ranges a variant of the `rms spectrum', which measures the absolute Root-Mean-Squared variability (rms) as a function of energy (e.g. see \citealt{Vaughan2003}).

Producing the rms spectrum involves allocating each photon event to a time and energy bin, dividing the light curve into segments consisting of $N$ time bins per segment and then working out the variance in each segment for each energy bin according to the following standard formula:
\begin{equation}
\sigma_{\rm xs}^2 = \frac{1}{N-1} \sum_{i=1}^N(X_i - \bar{X})^2 - \overline{\sigma_{\rm err}^2}
\end{equation}
where $X_i$ is the count rate in the $i^{\rm th}$ bin and $\bar{X}$ is the mean count rate in the segment. The expectation of Poisson noise variance, given by the average squared-error $\overline{\sigma_{\rm err}^2}$, is  subtracted, leaving the `excess variance', $\sigma_{\rm xs}^2$, in each segment. The excess variances can then be averaged over all of the segments of the light curve. The square root of the average excess variance plotted against energy forms the rms spectrum\footnote{This is absolute rms, since we are not normalising by the mean count rate.}.  By selecting the time bin size and the segment size, we can isolate different time-scales of variability, effectively replicating the Fourier resolved approach for the two time-scale ranges that we are interested in.  For this purpose, we choose two combinations of bin size and segment size.  To look at variations on shorter time-scales we choose 0.1~s time bins measured in segments of 4~s (i.e., 40 bins long), i.e., covering the frequency range 0.25-5~Hz\footnote{The upper frequency limit is set by the Nyquist frequency, $\nu_{\rm Nyq}=1/(2\Delta t)$ where $\Delta t$ is the time bin size.}.  For longer time-scale regions we use 2.7~s bins in segments of 270~s, i.e., covering the range 0.0037-0.185~Hz.  Note that these two frequency ranges do not overlap and also cover the two parts of both source PSDs which show distinct behaviour in soft and hard bands.

There is, however, a problem with the rms spectrum when signal to noise is low, e.g. at higher energies. It is possible for the expectation value of the Poisson variance term to be larger than the measured average variance term, producing negative average excess variances. If this is the case, it is not possible to calculate the rms at these energies and this introduces a bias towards the statistically higher-than-average realisations of rms values, which can still be recorded. In order to overcome these problems we have developed a technique called the `covariance spectrum'. The covariance is calculated according to the formula:
\begin{equation}
\sigma_{\rm cov}^2 = \frac{1}{N-1} \sum_{i=1}^N(X_i - \bar{X})(Y_i - \bar{Y})
\end{equation}
where $Y_i$ now refers to the light curve for a `reference band' running over some energy range where the variability signal-to-noise is large.  In this work, we use reference bands of 1--4 keV for EPIC-pn data and 3--5~keV for PCA data.  In other words, the covariance spectrum is to the rms spectrum what the cross-correlation function of a time series is to its auto-correlation function.  The covariance spectrum therefore does not suffer from the same problems as the rms spectrum, as no Poisson error term has to be subtracted, since uncorrelated noise tends to cancel out and any negative residuals do not affect the calculation. To remove the reference band component of the covariance, and produce a spectrum in count-rate units, we obtain the normalised covariance for each channel using:
\begin{equation}
\sigma_{\rm cov, norm} = \frac{\sigma_{\rm cov}^2}{\sqrt{\sigma_{\rm xs,y}^2}}
\end{equation}
where $\sigma_{\rm xs,y}^2$ is the excess variance of the reference band. Therefore, the only requirement for there being a valid, unbiased value of covariance at a given energy is that the reference excess variance is not negative. This is usually the case, since the reference band is chosen to include those energies with the largest absolute variability. When the covariance is being calculated for an energy channel inside the reference band, the channel of interest is removed from the reference band. The reasoning behind this is that if the channel of interest is duplicated in the reference band, the Poisson error contribution for that channel will not cancel and will contaminate the covariance. One can think of the covariance technique as applying a matched filter to the data, where the variations in the good signal-to-noise reference band pick out much weaker correlated variations in the energy channel of interest that are buried in noise. In this way the covariance spectrum picks out the components of the energy channel of interest that are correlated with those in the reference band.

\begin{figure}
\centering
\includegraphics[width=84mm]{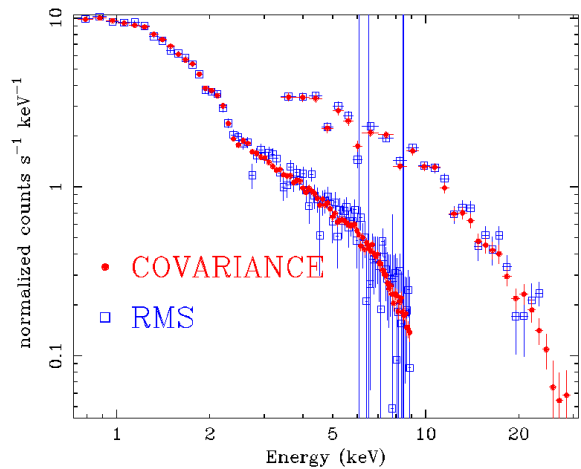}
\caption{Overlaid covariance and rms spectra for EPIC-pn and PCA data for SWIFT J1753.5-0127. The large error bars in the EPIC-pn data indicate negative excess variances where no rms value could be determined.}
\label{fig:covrms}
\end{figure}

It is important to note that the covariance spectrum only picks out the correlated variability component and is therefore a more appropriate measure than the rms spectrum in constraining the reprocessing of hard photons to soft photons, which will result in correlated variations. When the raw counts rms and covariance spectra are overlaid, as in figure~\ref{fig:covrms}, they match closely indicating that the reference band is well correlated with all other energies (the spectral `coherence' is high, e.g. see \citealt{Vaughan1997}). Another advantage of the matched filter aspect of the covariance spectrum is that it leads to smaller statistical errors than the rms spectrum. Specifically, the errors are given by:
\[
Err\left[\sigma_{\rm cov, norm}\right]=\sqrt{\frac{\sigma_{\rm xs,x}^2 \overline{\sigma_{\rm err,y}^2}+\sigma_{\rm xs,y}^2 \overline{\sigma_{\rm err,x}^2}+\overline{\sigma_{\rm err,x}^2}~\overline{\sigma_{\rm err,y}^2}}{NM\sigma_{\rm xs,y}^2}}
\]
where $M$ denotes the number of segments and subscripts x and y identify excess variances and Poisson variance terms for the channel of interest and reference band respectively.  This error equation can be derived simply from the Bartlett formula for the error on the zero-lag cross-correlation function, assuming that the source light curves have unity intrinsic coherence \citep{Bartlett1955,Box1976}. By comparison with equation B2 of \citealt{Vaughan2003}, using the relation:
\[
 \frac{\overline{\sigma_{\rm err,y}^2}}{\sigma_{\rm xs,y}^2} << \frac{\overline{\sigma_{\rm err,x}^2}}{\sigma_{\rm xs,x}^2}
\]
(which is true because the reference band has good signal to noise) it can be shown that the errors on the covariance are smaller than corresponding errors on the rms values.

Finally, we note that, since the covariance is analogous to the zero-lag cross-correlation function, it could be affected by intrinsic time lags in the data. Using measurements of the cross-spectral phase-lags between various energy bands, we have confirmed that the lags between hard and soft band variations are smaller than the time bin sizes used to make the long and short time-scale covariance spectra.  Thus, intrinsic time lags will have no effect on our results.

\subsection{SWIFT J1753.5-0127}
\begin{figure}
\centering
\includegraphics[width=60mm,angle=-90]{swiftj1753eeufspec.ps} 
\includegraphics[width=94mm]{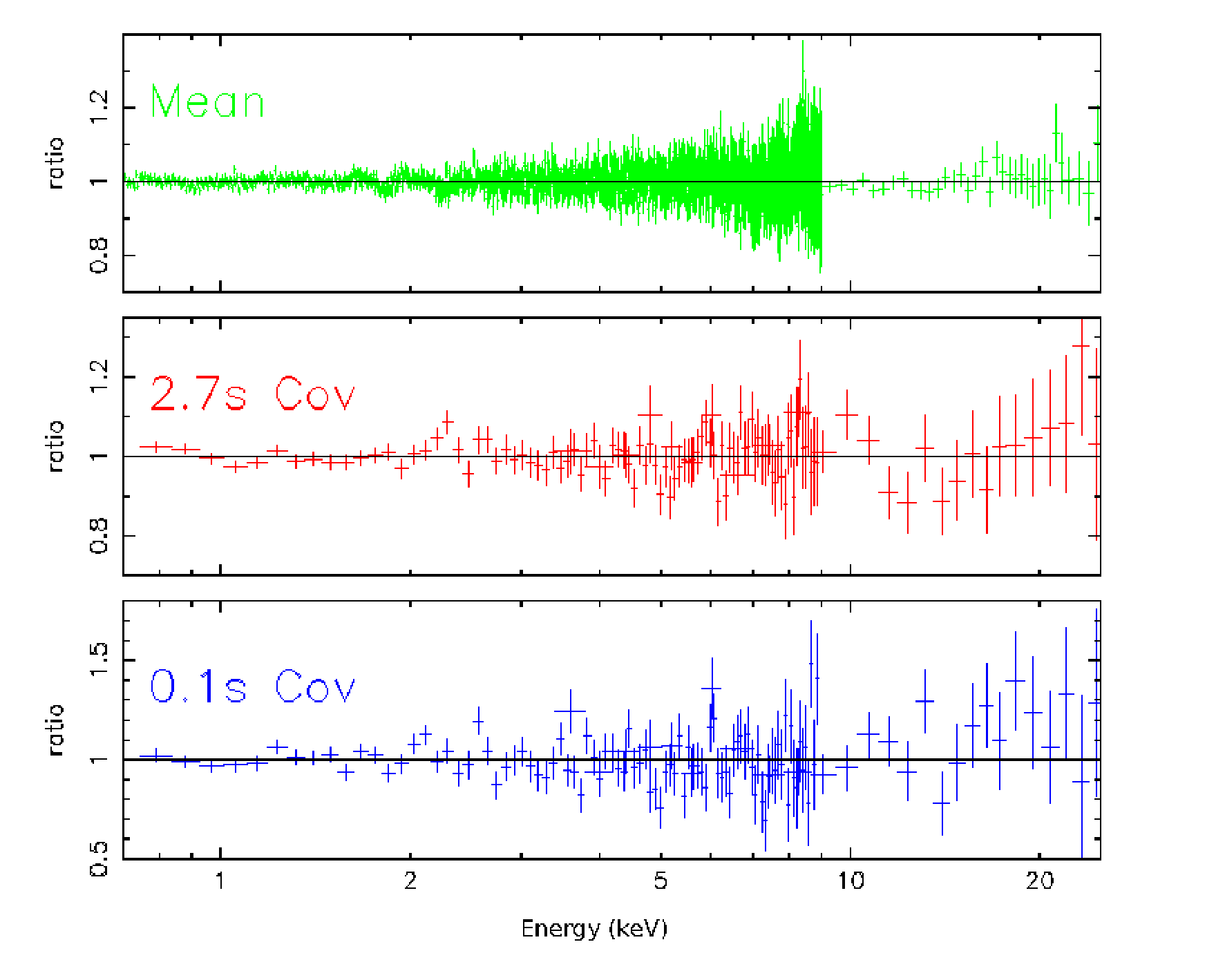}
\caption{Top panel: Unfolded spectra for SWIFT~J1753.5-0127 showing the relative contributions of the disc blackbody and power law components in the mean spectrum and short and long-time-scale covariance spectra. Dotted lines indicate spectral components of the short time-scale covariance, dashed lines indicate those of the long time-scale covariance. Bottom panel: SWIFT~J1753.5-0127 data/model ratios for the best-fitting disc blackbody plus power-law model.}
\label{fig:swifteeufspec}
\end{figure}

We first consider the covariance spectra of SWIFT~J1753.5-0127, and use {\sc xspec~v12}\footnote{http://heasarc.gsfc.nasa.gov/xanadu/xspec/} \citep{Arnaud1996} to fit the long and short-time-scale data together with the mean spectrum in order to identify the origin of the additional long-time-scale variability which can be seen in the soft band PSD.  The covariance spectra do not show sufficient signal-to-noise to detect the rather weak iron line present in the mean spectrum of this source \citep{Hiemstra2009,Reis2009}, so for simplicity we fit only a simple power-law and multicolour disc blackbody \textsc{diskbb}, together with neutral absorption. We fit the short and long-time-scale covariance and mean spectra simultaneously, tying the absorbing column density and renormalising constant for PCA data to be the same for the mean and covariance spectra. An F-test showed that the disk blackbody temperature does not change significantly between spectra, so that was also tied to be the same for all spectra.  The remaining parameters were allowed to be free between the covariance and mean spectra.  For those parameters that were free to vary, EPIC-pn and PCA values were tied together.  The $\chi^2$ of the final fit was 1802 for 1948 degrees of freedom (d.o.f.) and full fit parameters are listed in Table~\ref{tab:swift}.  The best-fitting unfolded spectra and the corresponding data/model ratios are shown in figure~\ref{fig:swifteeufspec}.

To interpret the fits to the covariance spectra, consider the case where the observed PSDs have identical shapes in both hard and soft bands (they may have different normalisations).  In this case, the ratio of soft to hard-band variability amplitudes measured over the same time-scale range will be identical for any given time-scale range.  Therefore, since (for high-coherence variations) the covariance spectrum quantifies variability amplitude as a function of energy, the shape of the covariance spectrum will be independent of time-scale.  On the other hand, if the soft band contains more long-time-scale variability relative to short-time-scale variability than the hard band, the long-time-scale covariance spectrum will appear {\it softer} than the short-time-scale covariance spectrum: lower energies show correspondingly greater variability on long time-scales and therefore larger fluxes in the covariance spectra.  Just such an effect is seen in the model fits to the covariance spectra: the long-time-scale covariance spectrum is softer than the short-time-scale covariance spectrum, because the disc blackbody normalisation is higher on long time-scales.  The power-law slope is remarkably similar on both long and short time-scales however.  Therefore the additional long-term variability in the soft-band PSD seems to result from {\it additional} variability of the disc blackbody, not any extra power-law variability (e.g. due to spectral pivoting).
\begin{table}
\centering
\caption{Fit parameters for mean and covariance spectra of SWIFT J1753.5-0127 for the model {\sc constant*phabs*(diskbb+powerlaw)}}
\begin{tabular}{l c c c}
\hline
Parameter & Mean & 2.7s-270s Cov & 0.1s-4s Cov \\
\hline
$C_{\rm PCA}$ & $1.222^{+0.006}_{-0.006}$ & tied to mean & tied to mean \\ [1ex]
$N_{\rm H}$ & $0.194^{+0.007}_{-0.006}$ & tied to mean & tied to mean \\[1ex]
$\Gamma$ & $1.625^{+0.005}_{-0.005}$ & $1.55\pm0.02$ & $1.55\pm0.03$ \\[1ex]
$A_{\rm pl}$ & $0.0731^{+0.0009}_{-0.0009}$ & $0.0118^{+0.0004}_{-0.0004}$ & $0.0128^{+0.0006}_{-0.0006}$ \\[1ex]
$kT_{\rm in}$ & $0.286^{+0.015}_{-0.014}$ & tied to mean & tied to mean \\[1ex]
$A_{\rm disc}$ & $320\pm25$ & $129\pm11$ & $75\pm20$ \\[1ex]
\hline
\end{tabular}
\begin{flushleft} From top to bottom, the parameters are: constant renormalising factor applied to model fit to PCA data; neutral absorbing column density ($10^{22}$~cm$^{-2}$); power-law photon index; power-law normalisation (photons cm$^{-2}$~s$^{-1}$~keV$^{-1}$ at 1 keV); disc blackbody inner radius temperature (keV); disc blackbody normalisation ($\left[r_{\rm in}/(D/10~{\rm kpc})\right]^{2}$ where $r_{\rm in}$ is the disc inner radius in km and $D$ is the distance in kpc).  All errors are 90~per cent confidence limits.  Since disc temperature and normalisation are highly correlated, the errors on blackbody normalisation are obtained while fixing temperature at the best-fitting value.
\end{flushleft}
\label{tab:swift}
\end{table}

\begin{figure}
\centering
\includegraphics[width=84mm]{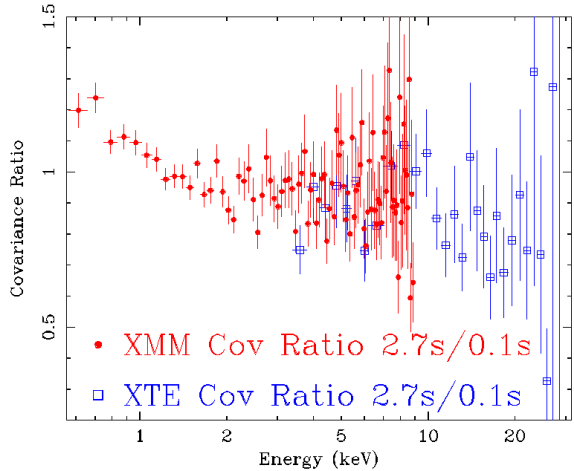}
\caption{Covariance ratios for SWIFT~J1753.5-0127 showing the extra blackbody component which contributes on longer timescales.}
\label{fig:swiftcovratios}
\end{figure}

We can show in a model-independent way that the blackbody component is the source of the additional long-term variability seen in the soft-band PSD, by measuring the `covariance ratio', i.e., plotting the ratio of long-time-scale to short time-scale covariance spectra.  By plotting the ratio, we remove most of the effects of the spectral response (except the smearing by the instrumental resolution) and can see more clearly the nature of any spectral differences.
The covariance ratio plot is shown in figure~\ref{fig:swiftcovratios}, and shows clearly that the excess in long-time-scale covariance arises abruptly below 2 keV, as expected if caused only by the disc emission, rather than being due to a gradual change across the whole spectrum which might be caused by differences in the power-law index. The interpretation of the short time-scale covariance spectrum is addressed later in section 4.4.

Having established that the reference band is well correlated with all other energies, the normalisations of the covariance spectral components, given in Table~\ref{tab:swift}, relative to those of the mean spectrum indicate the rms variability amplitude.  For example, dividing the disc blackbody covariance normalisations by that of the mean spectrum indicates fractional rms for the disc emission of 40~per cent and 23~per cent over long and short time-scale ranges respectively.  The power-law variability is more complicated to quantify, since the covariance spectra show significantly harder power-law slopes than the mean.  This result implies that the power-law fractional rms increases with energy.  The power-law normalisations imply that at 1~keV the fractional rms of the power-law is 16~per cent and 18~per cent over the long and short time-scale ranges respectively.

\subsection{GX 339-4}
\begin{figure}
\centering
\includegraphics[width=60mm,angle=-90]{gx339eeufspec.ps}
\includegraphics[width=94mm]{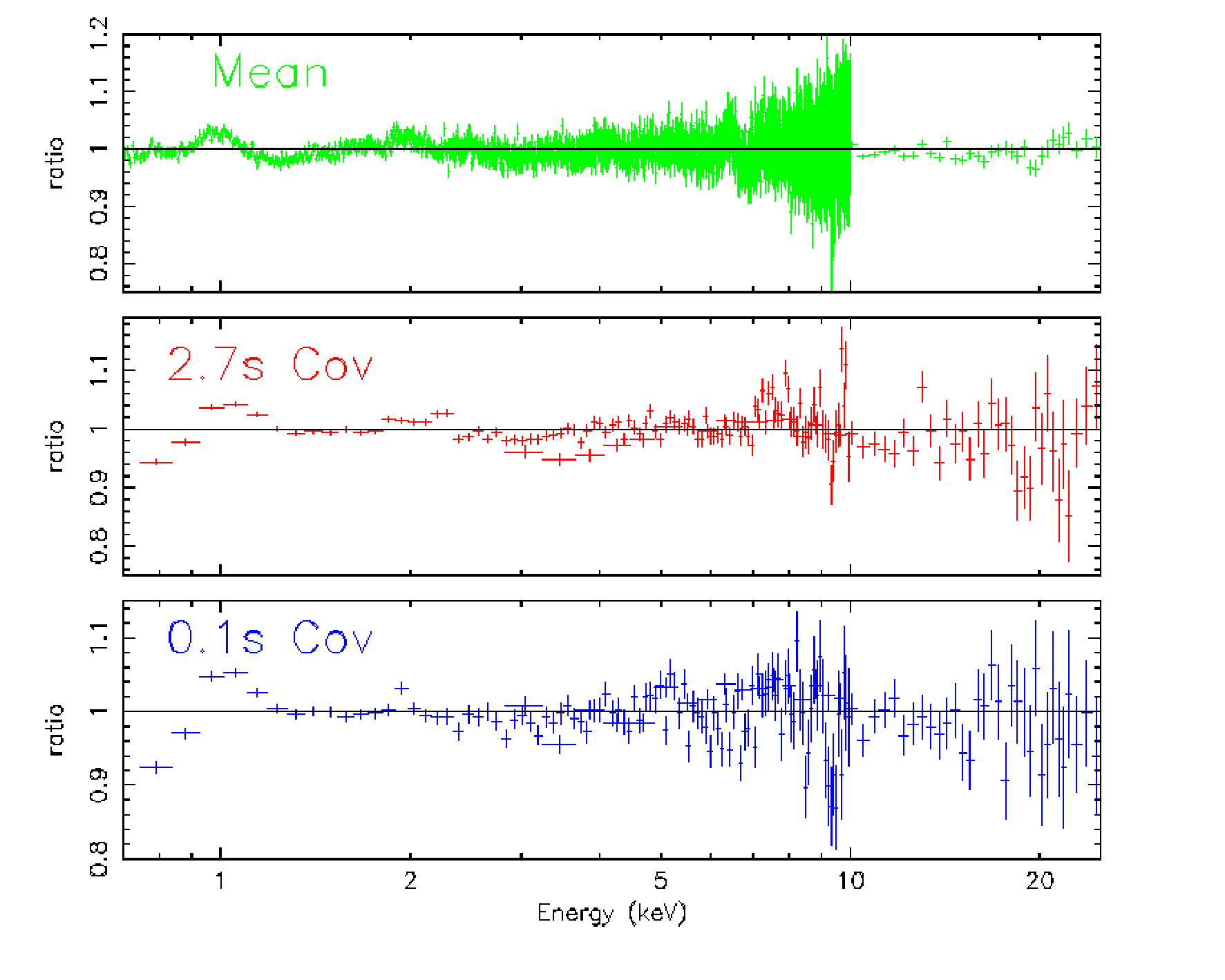}
\caption{Top panel: Unfolded spectra fitted for GX~339-4 showing the relative contributions of the blackbody, power law (incorporating reflection continuum) and iron line components. Dotted lines indicate spectral components of the short time-scale covariance, dashed lines indicate those of the long time-scale covariance. Bottom panel: GX~339-4 data/model ratios for the best-fitting disc blackbody plus power-law and relativistically smeared reflection model.}
\label{fig:gx339eeufspec}
\end{figure}

We next produced mean, long and short time-scale covariance spectra for EPIC-pn timing mode and PCA observations of GX 339-4.
We carried out fits to the spectra using a model consisting of a narrow Gaussian emission line (width $\sigma=0.01$~keV) and power law (including the multiplicative model \textsc{hrefl} to account for the reflected continuum from a cold optically thick disk), smeared with the convolution model \textsc{kdblur} in order to account for relativistic broadening of the reflected emission from material orbiting a maximally spinning black hole (see \citealt{Reis2008} for evidence for a high black hole spin in GX~339-4).  A multicolour disc model \textsc{diskbb} was also included to account for the soft X-ray excess. Strictly speaking this component should also be smeared, but it is not appropriate to smear this model with \textsc{kdblur}, as the disc temperature is itself a function of radius.  Since in the \textsc{kdblur} model the disc innermost radius and iron line radial emissivity are somewhat degenerate, the emissivity index was fixed at -3 (i.e., disc-like). Also, the outer radius was fixed at 400~$R_{\rm G}$. The remaining parameters, except the black body normalisation, reflection covering fraction, photon index, power law normalisation and Gaussian normalisation were tied to be the same between the mean and both covariance spectra (we confirmed with F-tests that the tied parameters do not vary significantly between the spectra).  As with SWIFT~J1753.5-0127, the EPIC-pn and PCA parameters were tied together for the same spectra and a renormalising constant factor was included to allow for differences in EPIC-pn and PCA flux calibration. The final fit had a $\chi^2$ of 2489 for 2199 d.o.f. and a full list of model parameters is given in Table~\ref{tab:gx}.  The best-fitting unfolded spectra and data/model ratios can be seen in figure~\ref{fig:gx339eeufspec}.
\begin{table}
 \centering
  \caption{Fit parameters for mean and covariance spectra of GX 339-4 for the model {\sc constant * phabs * (diskbb+kdblur(hrefl * powerlaw+gaussian))}}
  \begin{tabular}{l c c c}
  \hline
   Parameter & Mean & 2.7s-270s Cov & 0.1s-4s Cov \\
   \hline
$C_{\rm PCA}$ & $1.296\pm0.003$ & tied to mean & tied to mean \\[1ex]
$N_{\rm H}$ & $0.574\pm0.006$ & tied to mean & tied to mean \\[1ex]
$\Gamma$ & $1.766\pm0.006$ & $1.896\pm0.010$ & $1.875\pm0.011$ \\[1ex]
$A_{\rm pl}$ & $0.266\pm0.002$ & $0.110\pm0.001$ & $0.107\pm0.001$ \\[1ex]
$kT_{\rm in}$ & $0.177\pm0.001$ & tied to mean & tied to mean \\[1ex]
$A_{\rm disc}/10^{4}$ & $8.20\pm0.65$ & $3.20\pm0.26$ &  $2.46\pm0.21$\\[1ex]
$R_{\rm in}$ & $4.08\pm0.16$ & tied to mean & tied to mean \\[1ex]
$\theta_{\rm inc}$ & $40.2\pm1.4$ & tied to mean & tied to mean \\[1ex]
$E_{\rm Fe}$ & $6.42\pm0.07$ & tied to mean & tied to mean \\[1ex]
$A_{\rm Fe}/10^{-3}$ & $1.17\pm0.07$ & $0.61\pm0.07$ & $0.58\pm0.09$ \\[1ex]
$CF$ & $0.91\pm0.04$ & $1.13\pm0.08$ & $1.16\pm0.09$ \\[1ex]
\hline
\end{tabular}
\begin{flushleft}
For the definitions of the parameters listed in the first six rows see Table~\ref{tab:swift}. The additional parameters shown are for the relativistically smeared reflection (from top to bottom): disc innermost radius (units of $R_{\rm G}$); disc inclination (fixed to be the same in both {\sc kdblur} and {\sc hrefl}); Gaussian line energy; Gaussian normalisation (photons~cm$^{-2}$~s$^{-1}$); covering fraction of the reflection (where 1.0 corresponds to $2\pi$ steradians).
\end{flushleft}
\label{tab:gx}
\end{table}

\begin{figure}
\centering
\includegraphics[width=84mm]{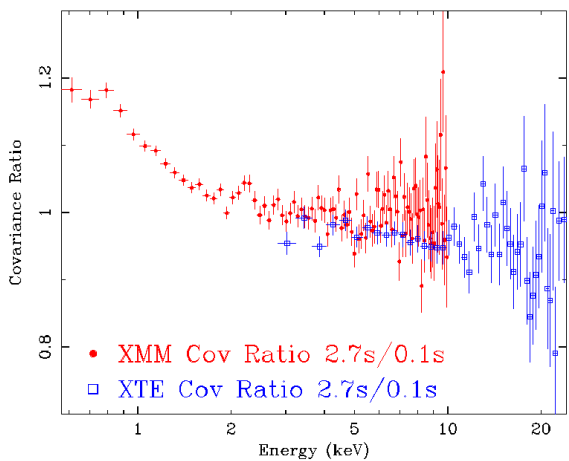}
\caption{Covariance ratios for GX 339-4 showing the extra thermal component on longer timescales}
\label{fig:gxcovratios}
\end{figure}

The fits to the covariance spectra for GX~339-4 show a similar pattern to that seen for SWIFT~J1753.5-0127, in that the long time-scale covariance spectrum is softer than the short time-scale covariance spectrum because of a significantly stronger disc blackbody component, while their power-law indices are very similar.  We can confirm this interpretation using a plot of the covariance ratio, which is shown in figure~\ref{fig:gxcovratios}, and shows a similar rise at low energies to that seen for SWIFT~J1753.5-0127, which underlines the interpretation that the disc blackbody component is the main contributor to the additional variability seen at low frequencies in the soft band in GX~339-4. The reader is referred to sections 4.1 and 4.4 for a consistent interpretation of the short time-scale covariance spectrum in GX~339-4 and SWIFT~J1753.5-0127. Note that the apparent emission feature around 2~keV is probably an instrumental effect, possibly related to mild pile-up in timing mode, since it disappears when larger regions of RAWX are excised from the data.  Due to the 1~per cent systematic included in the spectral fitting, this feature has no effect on the fit results.

Based on the ratio of component normalisations to those in the mean spectrum, the fractional rms values for the disc component are 39~per cent and 30~per cent over the long and short time-scale ranges respectively.  The corresponding power-law fractional rms values at 1~keV are 41~per cent and 40~per cent.  Using the same approach, we can also define a fractional rms for the iron line emission for long and short time-scales, at $\sim50$~per cent in each case.  Interestingly, the $\sim25$~per cent increase in iron line rms over that of the power-law emission (which drives the line variability) is comparable to the increase in reflection covering fraction from the mean to covariance spectra.  These results may imply the presence of an additional constant power-law component which dilutes the power-law fractional variability but does not contribute to reflection.  It is also interesting to note that the power-law component in the covariance spectra for GX~339-4 is softer than in the mean spectrum, i.e., the difference is in the opposite sense to that seen in SWIFT~J1753.5-0127.

\section{Discussion}
\label{discuss}
We have shown that disc blackbody emission contributes significantly to the X-ray variability spectra in the hard state of the black hole candidates SWIFT~J1753.5-0127 and GX~339-4, and moreover, that the disc emission is the origin of the additional soft band variability seen on longer time-scales in both sources, which manifests itself as an enhanced low-frequency component in the PSD.  In this section, we discuss the evidence for a connection between disc and power-law variability through X-ray reprocessing, and then consider two possible explanations for the enhanced disc variability on long time-scales, in terms of geometry changes or fluctuations intrinsic to the accretion disc.  Finally we will compare our results and interpretation to the wider picture of different accretion states.

\subsection{Power-law variability and evidence for thermal reprocessing}
The power-law emission clearly dominates the X-ray luminosity in both sources, as can be seen in figures~\ref{fig:swifteeufspec} and \ref{fig:gx339eeufspec}.  In this situation, if the disc sees a reasonable fraction of the power-law emission, we should expect that X-ray heating of the disc, i.e., thermal reprocessing of power-law emission, will produce a significant fraction of the observed disc luminosity.  In fact, the thermal reprocessed emission is directly related to the disc reflection component in the spectrum: if a fraction of incident power-law luminosity $f$ is reflected by the disc (both through Compton reflection and emission line fluorescence), then a fraction $1-f$ must be absorbed and will be reprocessed into thermal blackbody radiation.  The reflected fraction $f$ depends on disc ionisation state but simple exploration of the disc reflection models in {\sc xspec} shows that it is typically 30-40~per cent of the incident luminosity, so that around 60-70~per~cent of the incident luminosity is reprocessed into disc blackbody emission.  For typical iron K$\alpha$ line equivalent widths (around 1~keV with respect to the reflection continuum), we then expect the line flux to be of order 1~per cent of the thermally reprocessed flux.  We might reasonably assume that the blackbody component in the short time-scale covariance spectrum is produced by thermal reprocessing of the varying power-law, which also drives the line emission in the same spectrum. The unabsorbed disc flux is $\sim4.5\times 10^{-10}$~erg~cm$^{-2}$~s$^{-1}$, and the iron line flux is $\sim6\times10^{-12}$~erg~cm$^{-2}$~s$^{-1}$ which is 1.3~per cent of the reprocessed flux, i.e., consistent with a reprocessing origin for the thermal emission, at least on short time-scales.

In SWIFT~J1753.5-0127, the disc blackbody emission is considerably weaker than in GX~339-4, and correspondingly we would expect a relatively weak iron line, with a few tens of eV equivalent width, which is only just consistent with the lower-limits on line strengths reported by \citet{Hiemstra2009} and \citet{Reis2009}.  It is possible that the disc in SWIFT~J1753.5-0127 is substantially ionised (e.g. see \citealt{Reis2009}) which would enhance line emission relative to the absorbed (and hence thermally reprocessed) emission.  

Since it is likely that there is substantial X-ray heating of the disc, one must interpret the observed blackbody normalisations in the mean spectra with caution\footnote{The absolute normalisation in the covariance spectra is a function of variability amplitude and cannot be interpreted in the same way as in the mean spectrum.}.  The disc emissivity may be more centrally concentrated than the theoretically expected $R^{-3}$ law, and so the normalisations indicate better the emitting surface area and cannot be simply translated to an inner radius.  Nonetheless, the inferred emitting areas are still relatively small, implying distance scales of tens to hundreds of km (assuming distances $>7$~kpc and 6-15~kpc for SWIFT~J1753.5-0127 and GX~339-4 respectively; \citealt{Zurita2008,Hynes2004}).  We also note here that although the need for disc blackbody emission to explain the spectrum of SWIFT~J1753.5-0127 has been questioned by \citet{Hiemstra2009}, the model-independent covariance ratio plots in figure~\ref{fig:swiftcovratios} show that a distinct soft component must be present in order to explain the difference in the shapes of the covariance spectra.

It is interesting to note that the power-law indices of the covariance spectra are different to those of the mean spectra, but in an opposite sense for each of the two sources considered here: compared to the mean spectrum SWIFT~J1753.5-0127 shows a harder power-law in the covariance spectra, while GX~339-4 shows a softer power-law.  The difference may be caused by flux-dependent spectral pivoting or steepening, so that as flux increases the spectrum gets softer in GX~339-4, increasing the covariance at soft energies relative to the mean, while the opposite effect occurs in SWIFT~J1753.5-0127 (it hardens as it gets brighter).  The difference in behaviour may be related to the source luminosity: if they lie at similar distances SWIFT~J1753.5-0127 is at least a factor 10 less luminous than GX~339-4, implying a significantly lower accretion rate.  Correlations between flux and spectral-hardness have been seen in BHXRB hard states, and interestingly the sign of the correlation appears to switch over from negative to positive at low luminosities, both on short time-scales \citep{Axelsson2008} and in the long-term global correlation \citep{Wu2008}.  The same switch in flux-hardness correlation could be related to the different power-law behaviour of SWIFT~J1753.5-0127 and GX~339 which we see here.  

The same pattern appears on both long and short time-scales, which show almost identical power-law indices in their covariance spectra, so that the effect of the power-law spectral variability will be to change the normalisation of the PSD, but not the shape.  Since most of the power-law luminosity will be found at tens of keV (assuming a thermal cutoff at around 100~keV), spectral steepening with flux will cause the observed GX~339-4 0.5-10~keV variability amplitude to be enhanced compared to the true luminosity variations. Conversely, observed variations in SWIFT~J1753.5-0127 will be smaller than the total luminosity variations.  Therefore, the fractional luminosity variations for SWIFT~J1753.5-0127 and GX~339-4 may be similar, but GX~339-4 shows a significantly greater normalisation in the PSDs shown in figure~\ref{psds}. 

\subsection{Variable coronal geometry}
We have seen that it is likely that reprocessing of the power-law drives at least some of the disc variability seen in hard state sources, and possibly all of it on time-scales $<1$~s. However, model-independent covariance ratio plots and spectral fitting show that the covariance spectra of both sources demonstrate increased disc blackbody variability with respect to the power law on longer timescales.   There are several possible explanations for this pattern, but the key thing that any successful model needs to achieve is an increase in disk variability on longer timescales without a concomitant rise in power law variability. Also, it is important to note that the additional disc variability must still be correlated with power-law variations, because the enhanced disc variations appear in the covariance spectrum, which is identical to the rms spectrum over the high signal-to-noise energy range covered by the disc (i.e., coherence is unity). If the disc variations were independent of the power-law they would cancel to some extent, since they would be uncorrelated with the power-law component, and covariance would be smaller than the rms.  Thus the blackbody variations map on to power-law variations but with larger amplitude.

One possibility is to change the geometry of the system on longer time-scales.  For example, a variable coronal scale height on longer timescales could lead to changes in the solid angle of disc heated by the power-law, thus increasing the disc blackbody variability amplitude relative to the power-law. Weaker correlated power-law variation could then be produced if scale-height correlates with power-law luminosity.  Alternatively, correlated power-law variations could be due to variable seed photon numbers from the disc due to the enhanced variable heating, but in either case the additional power-law variability must be of smaller amplitude than the observed blackbody variability.  Regardless of these model-dependent arguments, one can make a simple observational test of the variable-geometry model, by comparing the variability of reflection on long and short time-scales.

\begin{figure}
\centering
\includegraphics[width=84mm]{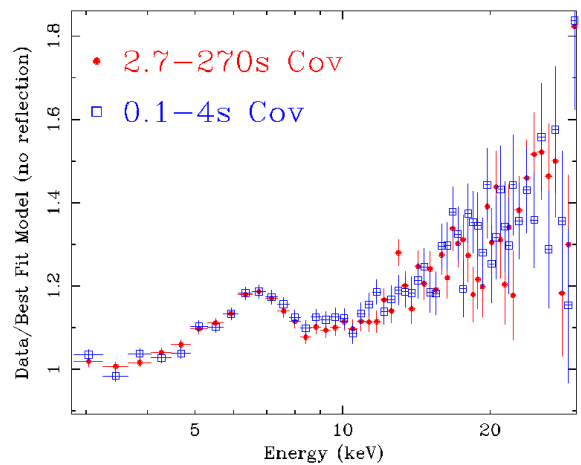}
\caption{Data to model ratios of covariance spectra in GX 339-4 with reflection normalisations set to zero}
\label{fig:gxrefl}
\end{figure}

GX~339-4 shows significant reflection features in its covariance spectra, so any variation in coronal geometry should manifest itself as increased variability in the reflection components as the disk sees the varying power law.  However, the spectral fit parameters given in Table~\ref{tab:gx} indicate that there is little change in reflection amplitude between long and short time-scales, both in terms of reflection covering fraction and iron line equivalent width (e.g. ratio of line flux to power-law normalisation, which is meaningful because the power-law indices are so similar). To highlight this similarity in the covariance spectra, we show in figure~\ref{fig:gxrefl} the data/model ratios for the PCA spectra with the reflection components taken out of the fit.  The spectra demonstrate no significant change in the reflection continuum and associated iron line. This implies that if there is any change in geometry, it is small, and the increase in the thermal component of the variability spectra on longer timescales has a different origin. 

\begin{figure}
\centering
\includegraphics[width=60mm,angle=-90]{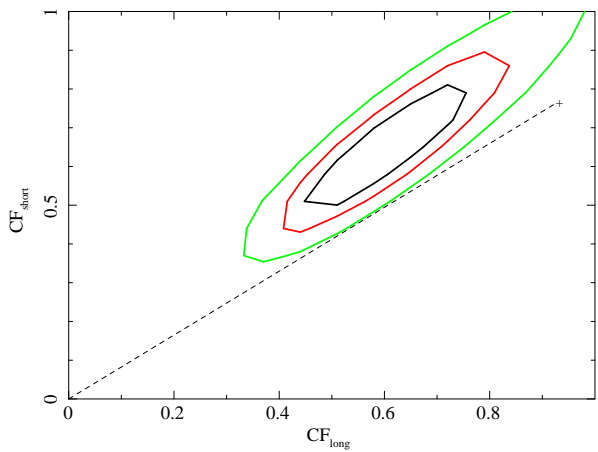}
\caption{GX 339-4 confidence contour plots for the permitted range of reflection covering fraction in simultaneous fits to the long and short-time-scale covariance spectra. The dotted line passes through the 99~per cent confidence contour, so its gradient defines an upper limit to the ratio of long to short-time-scale covering fraction. Also shown are the 90 and 68~per cent confidence level contours.}
\label{fig:contour}
\end{figure}

To place this result on a more rigorous statistical footing, we show in
figure~\ref{fig:contour} a contour plot of the best-fitting short and long-time-scale reflection covering fractions, which are obtained only from fits to the PCA data, which are most sensitive to the reflection continuum\footnote{As a result, the best-fitting covering fraction is significantly lower than observed in the joint fit with EPIC-pn, but since we are interested in the {\it relative} change in reflection between different time-scales, this does not affect our basic result.}. If variable geometry is the cause of the enhanced long-term blackbody variability, we would expect the covering fraction to show a similar enhancement in variability on long time-scales.  The dashed line in the figure shows the largest ratio of short-to-long-time-scale covering fraction which crosses the 99~per cent confidence contour, with a value of 0.81, placing a 99~per cent confidence upper limit of 23~per cent on any increase in the covering fraction on long time-scales compared to short time-scales (the 90~per cent confidence upper limit on any increase is 12~per cent). In contrast, the long-time-scale blackbody variability amplitude increases by 30~per cent compared to that on shorter time-scales, which cannot be explained by the permitted increase in variable reflection at greater than the 99~per cent confidence level.  Thus, in GX~339-4 at least, we can rule out long-time-scale changes in coronal geometry as a viable explanation for the enhanced blackbody variability amplitude. Note also that the same arguments apply to any other non-geometric arguments which seek to produce the extra long-term disc variability by varying the power-law contribution as seen by the disc, e.g through variable beaming of the power-law towards the disc. Due to the weakness of the iron line in SWIFT J1753.5-0127 it is not possible to place similar constraints on coronal geometry changes in this source, although by analogy we expect the same interpretation to apply.

\subsection{Intrinsic disc variability}

Having established that variable coronal geometry cannot explain the extra blackbody variability on longer timescales, we are left to consider the possibility that the variations are intrinsic to the disc itself.  Perhaps the simplest possibility is that the fluctuations are due to accretion rate fluctuations in the disc which undergo viscous damping before they reach the corona.  Such damping is expected in thin discs (e.g. see \citealt{Churazov2001}), but will not be as significant in geometrically thick flows, which might correspond to the corona.  Therefore one can envisage that the long time-scale variability, e.g. corresponding to the low-frequency Lorentzian in the PSD, is generated in the thin disc, producing relatively large amplitude blackbody variations but being damped before reaching an inner coronal emitting region (perhaps inside the innermost radius of the thin disc). The correlated power-law variability could then arise either from the residual undamped variations in accretion rate which reach the corona, or be driven by seed photon variations from the disc.

If the inner radius of the disc were to fluctuate on long time-scales, then the varying disc area would also introduce extra blackbody variability. This model is in some sense analogous to that of changes in coronal scale height, since any change in disc area will vary the solid angle of disc seen by the corona which will cause an increase in reflection as well as correlated power-law variability driven by seed photon variations.  One could mitigate these effects if the coronal properties were linked to those of the disc inner-radius, e.g. as inner radius decreases, so does coronal scale-height.  This situation might be expected if the corona is formed by evaporation of the disc \citep{Meyer-Hofmeister2003}, so that condensation of the disc will drain and cool the coronal plasma, leading to a reduction in scale-height.

\subsection{Wider implications}
We have established that the hard-state variability on time-scales greater than seconds, corresponding to the low-frequency Lorentzian PSD component, is very probably produced by variations intrinsic to the accretion disc, perhaps in the form of propagating accretion rate fluctuations, as envisaged by \citet{Lyubarskii1997} to explain the broad shapes of observed PSDs.  These variations then manifest as weaker power-law variations, either through propagation of accretion fluctuations to the corona \citep{Churazov2001}, or through variations of seed photons from the disc, which are Compton upscattered in the corona.  On shorter time-scales, the blackbody variations are probably mostly produced by X-ray heating of the disc by the power-law, which is a required outcome of the X-ray reflection directly observed in these systems.  The exact split between direct emission from intrinsic disc variations and X-ray heating is difficult to judge, since some residual intrinsic variations may remain on short time-scales, and X-ray heating will also contribute on long time-scales.  But in the case of intrinsic disc accretion fluctuations, the variable X-ray heating is itself a product of those fluctuations, and so any intrinsic disc variations on long time-scales must be large, at least comparable to the fractional rms of blackbody emission, i.e., 40~per cent in both sources. 

It is interesting to contrast the large intrinsic disc variability in these hard state sources with that in the soft states, where the disc emission dominates the bolometric luminosity.  The soft states are well-known for showing very weak, if any, variability \citep{Homan2001}, and the strongest variability which is seen, e.g. in Cyg~X-1, is associated with the power-law, with the disc being remarkably constant \citep{Churazov2001}.  Therefore it seems likely that hard state discs are inherently unstable compared to soft state discs.  This difference may represent just another observable distinction between hard and soft states, but it is interesting to speculate that it may play a more primary role in creating the other observed differences, such as a strong corona and jet formation in the hard state.  Certainly, it seems likely that intrinsic disc variability plays an important role in determining the PSD shape in the hard state. For instance, the low energy PSD of GRS~1915+105 in the $\chi$-class hard intermediate state \citep{Rodriguez2004} shows a low frequency component which disappears at higher energies, possibly indicating that this source is demonstrating intrinsic disc variability.

It is worth noting here that X-ray/optical  studies of AGN also show evidence for reprocessing on short time-scales and intrinsic disc variations on longer time-scales \citep{Uttley2003,Arevalo2008,Breedt2009}.  However these AGN are relatively luminous and radio-quiet and so are likely analogues of BHXRB soft states, perhaps indicating that disc-stability shows a mass-dependence, e.g. related to the transition between gas- and radiation-pressure dominated discs \citep{Maccarone2003}.

Our results strongly suggest that disc variations are responsible for the low-frequency component in the hard state PSD.  A number of authors have suggested that the low-frequency Lorentzian corresponds to the viscous time-scale of the inner, truncation radius of the thin disc (e.g. \citealt{Churazov2001,Done2007}), a view which is consistent with our results.  The higher-frequency PSD components may then be produced in the corona, which is likely to be geometrically thick so will show naturally shorter variability time-scales.  The viscous time-scale scales with scale-height ($H$) over radius ($R$) as $(H/R)^{2}$.  Assuming the ratio of disc scale-height to radius $H/R\sim \alpha \sim 0.1$ (where $\alpha$ is the viscosity parameter), the predicted inner disc radius corresponding to the observed low-frequency PSD peak around $\sim 0.05$~Hz is 15~R$_{\rm G}$ (see \citealt{Done2007}).  However, this radius may be even smaller for smaller $H/R$, which would make the disc truncation radius consistent with the results from fits to the iron line (this work, and \citealt{Reis2008,Reis2009}).  However, in the latter case the corona would need to be very compact, or seed photon variations from intrinsic disc variability would modulate the power-law with a similar amplitude to the disc.  In either case, assuming the disc is thin and that it varies on the viscous time-scale, its inner radius must be relatively small.  Such a picture is very different from earlier models for the hard state, where the disc is highly truncated and power-law emission is produced by a very extended corona or ADAF (e.g. \citealt{Esin1997}).

\section{Conclusions}
\label{conc}

Spectral fits to observations of BHXRB sources in the hard state show increasing evidence for both power law and black body components. In this work we have explored the hard-state variability of two sources with known soft excesses, SWIFT~J1753.5-0127 and GX~339-4. Our findings are summarised below.

\begin{enumerate}[1.]
\item We have introduced a new spectral analysis technique, the covariance spectrum, which measures the correlated variability in different energy bands. This technique overcomes the problems of low signal-to-noise and bias associated with the rms spectrum and has smaller statistical errors.

\item PSDs of the two sources demonstrate larger low-frequency power in the soft band.

\item The longer time-scale (2.7-270~s) covariance spectra of both sources are softer than the short time-scale (0.1-4~s) covariance spectra, due to additional disc variability, i.e., extra disc variability occurs on longer time-scales without a concomitant rise in power law variability on such timescales. However, the coherence of the rms and covariance spectra show clearly that disc variations are not independent of the power law variations.

\item The strength of reflection features that are detected in the short time-scale covariance spectra of GX~339-4 are consistent with the observed blackbody variations on those time-scales being driven by thermal reprocessing of the power-law emission absorbed by the disc.  However,
the reflection covering fraction and iron line equivalent width show little change between short and long time-scales, implying that additional reprocessing, due to coronal geometry change, is not responsible for the additional blackbody variability seen on longer time-scales.

\item The extra blackbody variability seen on longer time-scales appears to be intrinsic to the accretion disc itself, giving rise to the extra low-frequency power in the PSD. This represents the first clear evidence that the low-frequency Lorentzian component in hard state PSDs is produced by disc variability.  Models invoking damped mass accretion rate variations or oscillations in the disc truncation radius can satisfactorily explain the observed pattern of variability. 

\item The implication of such variations occurring in a thin disc on viscous timescales, is that the disc truncation radius is $<20$~R$_{\rm G}$.

\end{enumerate}

This work highlights the importance of measuring spectral variability on a range of time-scales. Mean spectra, which describe the average properties of a source, provide no information on how different spectral components are related to one another as a function of time. By using the covariance spectra we have been able to disentangle the correlated spectral components in these two sources, identify thermal reprocessing as the mechanism by which variability is correlated in different bands, produce model-independent evidence for additional blackbody variability on longer time-scales and therefore associate intrinsic disc variability with the low frequency Lorentzian feature seen in hard-state PSDs. 

\section*{Acknowledgments}
We would like to thank Beike Hiemstra and the anonymous referee for useful comments. We are grateful to Maria D\'{i}az Trigo for providing the GX 339-4 EPIC-pn events files and helpful advice.  TW is supported by an STFC postgraduate studentship grant, and PU is supported by an STFC Advanced Fellowship.  This research has made use of data obtained from the High Energy Astrophysics Science Archive Research Center (HEASARC), provided by NASA's Goddard Space Flight Center, and also made use of NASA's Astrophysics Data System.

\label{lastpage}


\begin{thebibliography}{1}
\bibitem[\protect\citeauthoryear{Ar\'{e}valo et al.}{2008}]{Arevalo2008} Ar\'{e}valo, P., Uttley, P., Kaspi, S., Breedt, E., Lira, P., M$^{\rm c}$Hardy, I. M., 2008, MNRAS, 389, 1479
\bibitem[\protect\citeauthoryear{Arnaud}{1996}]{Arnaud1996} Arnaud, K.~A. 1996, Astronomical Data Analysis Software and Systems V, 101, 17 
\bibitem[\protect\citeauthoryear{Axelsson et al.}{2008}]{Axelsson2008} Axelsson, M., Hjalmarsdotter, L., Borgonovo, L., Larsson, S., 2008, A\&A, 490, 253
\bibitem[\protect\citeauthoryear{Bartlett}{1955}]{Bartlett1955} Bartlett, M.S., 1955, An Introduction to Stochastic Processes, CUP, Cambridge
\bibitem[\protect\citeauthoryear{Box \& Jenkins}{1976}]{Box1976} Box, G.E.P., Jenkins, G.M., 1976, Time Series Analysis: Forecasting and Control, 2nd edn. Holden-Day, San Francisco
\bibitem[\protect\citeauthoryear{Breedt et al.}{2009}]{Breedt2009} Breedt, E., et al., 2009, MNRAS in press (arXiv0812.0810) 
\bibitem[\protect\citeauthoryear{Churazov et al.}{2001}]{Churazov2001} Churazov, E., Gilfanov, M., \& Revnivtsev, M., 2001, MNRAS, 321, 759
\bibitem[\protect\citeauthoryear{Done et al.}{2007}]{Done2007} Done, C., Gierli{\'n}ski, M., \& Kubota, A., 2007, A\&ARv, 15, 1 
\bibitem[\protect\citeauthoryear{Esin, McClintock, \& Narayan}{1997}]{Esin1997} Esin A.~A., McClintock J.~E., Narayan R., 1997, ApJ, 489, 865
\bibitem[\protect\citeauthoryear{Gierli{\'n}ski et al.}{2008}]{Gierlinski2008} Gierli{\'n}ski, M., Done, C., Page, K., 2008, MNRAS, 388, 753
\bibitem[\protect\citeauthoryear{Gilfanov, Churazov \& Revnivtsev}{1999}]{Gilfanov1999} Gilfanov, M., Churazov, E., \& Revnivtsev, M., 1999, A\&A, 352, 182
\bibitem[\protect\citeauthoryear{Hiemstra et al.}{2009}]{Hiemstra2009} Hiemstra, B., Soleri, P., Mendez, M., Belloni, T., Mostafa, R., Wijnands, R., 2009, MNRAS in press (arXiv0901.2255)
\bibitem[\protect\citeauthoryear{Homan et al.}{2001}]{Homan2001} Homan, J., Wijnands, R., van der Klis, M., Belloni, T., van Paradijs, J., Klein-Wolt, M., Fender, R., \& M{\'e}ndez, M., 2001, ApJS, 132, 377
\bibitem[\protect\citeauthoryear{Hynes et al.}{2004}]{Hynes2004} Hynes, R. I., Steeghs, D., Casares, J., Charles, P. A., O'Brien, K., 2004, ApJ, 609, 317
\bibitem[\protect\citeauthoryear{Lyubarskii}{1997}]{Lyubarskii1997} Lyubarskii, Y.~E., 1997, MNRAS, 292, 679 
\bibitem[\protect\citeauthoryear{Maccarone,  Gallo \& Fender}{2003}]{Maccarone2003} Maccarone, T. J., Gallo, E., Fender, R., 2003, MNRAS, 345, L19
\bibitem[\protect\citeauthoryear{Malzac}{2007}]{Malzac2007} Malzac, J., 2007, APSS, 311, 149 
\bibitem[\protect\citeauthoryear{M$^{\rm c}$Hardy et al.}{2006}]{McHardy2006} M$^{\rm c}$Hardy, I. M., K\"{o}rding, E., Knigge, C., Uttley, P., Fender, R. P., 2006, Nature, 444, 730
\bibitem[\protect\citeauthoryear{Mayer \& Pringle}{2007}]{Mayer2007} Mayer, M., \& Pringle, J.~E., 2007, MNRAS, 376, 435 
\bibitem[\protect\citeauthoryear{Meyer-Hofmeister \& Meyer}{2003}]{Meyer-Hofmeister2003} Meyer-Hofmeister, E., Meyer, F. 2003, A\&A, 402, 1013
\bibitem[\protect\citeauthoryear{Miller et al.}{2006}]{Miller2006a} Miller J.~M., Homan J., Steeghs D., Rupen M., Hunstead R.~W., Wijnands R., Charles P.~A., Fabian A.~C., 2006, ApJ, 653, 525 
\bibitem[\protect\citeauthoryear{Miller, Homan, \& Miniutti}{2006}]{Miller2006b} Miller J.~M., Homan J., Miniutti G., 2006, ApJ, 652, L113 
\bibitem[\protect\citeauthoryear{Narayan \& Yi}{1994}]{Narayan1994} Narayan, R., \& Yi, I., 1994, ApJ, 428, L13 
\bibitem[\protect\citeauthoryear{Reis et al.}{2008}]{Reis2008} Reis, R. C., Fabian, A. C., Ross, R. R., Miniutti, G., Miller, J. M., Reynolds, C., 2008, MNRAS, 387, 1489
\bibitem[\protect\citeauthoryear{Reis et al.}{2009}]{Reis2009} Reis, R. C., Fabian, A. C., Ross, R. R., Miller, J. M., 2009, MNRAS in press (arXiv0902.1745)
\bibitem[\protect\citeauthoryear{Remillard \& McClintock}{2006}]{Remillard2006} Remillard R.~A., McClintock J.~E., 2006, ARA\&A, 44, 49 
\bibitem[\protect\citeauthoryear{Revnivtsev, Gilfanov \& Churazov}{1999}]{Revnivtsev1999} Revnivtsev, M., Gilfanov, M., \& Churazov, E., 1999, A\&A, 347, L23
\bibitem[\protect\citeauthoryear{Revnivtsev, Gilfanov \& Churazov}{2001}]{Revnivtsev2001} Revnivtsev, M., Gilfanov, M., \& Churazov, E., 2001, A\&A, 380, 520 
\bibitem[\protect\citeauthoryear{Rodriguez et al.}{2004}]{Rodriguez2004} Rodriguez, J., Corbel, S., Hannikainen, D.~C., Belloni, T., Paizis, A., \& Vilhu, O., 2004, ApJ, 615, 416 
\bibitem[\protect\citeauthoryear{Shakura \& Sunyaev}{1973}]{Shakura1973} Shakura N.~I., Sunyaev R.~A., 1973, A\&A, 24, 337 
\bibitem[\protect\citeauthoryear{Uttley et al.}{2003}]{Uttley2003} Uttley, P., Edelson, R., M$^{\rm c}$Hardy, I. M., Peterson, B. M., Markowitz, A., 2003, ApJ, 584, L53
\bibitem[\protect\citeauthoryear{Vaughan \& Nowak}{1997}]{Vaughan1997} Vaughan, B. A., Nowak, M. A., 1997, ApJ, 474, L43
\bibitem[\protect\citeauthoryear{Vaughan et al.}{2003}]{Vaughan2003} Vaughan, S., Edelson, R., Warwick, R. S., Uttley, P., 2003, MNRAS, 345, 1271
\bibitem[\protect\citeauthoryear{Witt, Czerny \& Zycki}{1997}]{Witt1997} Witt, H. J., Czerny, B., Zycki, P. T., 1997, MNRAS, 286, 848
\bibitem[\protect\citeauthoryear{Wu \& Gu}{2008}]{Wu2008} Wu, Q., Gu, M., 2008, ApJ, 682, 212
\bibitem[\protect\citeauthoryear{Zurita et al.}{2008}]{Zurita2008} Zurita, C., Durant, M., Torres, M. A. P., Shahbaz, T., Casares, J., Steeghs, D., 2008, ApJ, 681, 1458








\end{thebibliography}
\end{document}